%% file: main.tex
\pdfoutput=1
\documentclass[12pt,a4paper]{article}

\usepackage{ifthen} 
\newboolean{pdflatex}
\setboolean{pdflatex}{true} 

\newboolean{articletitles}
\setboolean{articletitles}{true} 

\newboolean{uprightparticles}
\setboolean{uprightparticles}{false} 

\def\paperauthors{LHCb collaboration} 
\def\paperasciititle{Angular analysis of the B+ -> pi+ mu+ mu- decay} 
\def\papertitle{Angular analysis of the $B^+\to\pi^+\mu^+\mu^-$ decay}
\def\paperkeywords{{High Energy Physics}, {LHCb}} 
\def\papercopyright{\the\year\ CERN for the benefit of the LHCb collaboration} 
\def\paperlicence{CC BY 4.0 licence}
\def\paperlicenceurl{https://creativecommons.org/licenses/by/4.0/}

\input{preamble}
\usepackage{longtable} 
\usepackage{booktabs}

\def\BuJpsiK      {\decay{\Bp}{\jpsi\Kp}}
\def\BuJpsiPi     {\decay{\Bp}{\jpsi\pip}}
\def\JpsiMuMu     {\decay{\jpsi}{\mumu}} 
\def\BuPiMuMu     {\decay{\Bp}{\pip\mumu}}
\def\BcPiMuMu     {\decay{\Bc}{\pip\mumu}} 
\def\BuKMuMu      {\decay{\Bp}{\Kp\mumu}}
\def\BuPiPiPi     {\decay{\Bp}{\pip\pim\pip}}
\def\BuKPiPi      {\decay{\Bp}{\Kp\pim\pip}}
\def\BsfMuMu      {\decay{\Bs}{f_0(980)\mumu}}
 
\def\PiMuMu       {\ensuremath{\pip\mumu}\xspace} 
\def\AFB          {\ensuremath{A_{\rm FB}}\xspace}
\def\FH           {\ensuremath{F_{H}}\xspace}
\def\Thetal     {\ensuremath{\theta_l}\xspace}
\def\CosThetal     {\ensuremath{\cos\theta_l}\xspace}

\begin{document}

\renewcommand{\thefootnote}{\fnsymbol{footnote}}
\setcounter{footnote}{1}

\input{title-LHCb-PAPER}


\renewcommand{\thefootnote}{\arabic{footnote}}
\setcounter{footnote}{0}

\cleardoublepage


\pagestyle{plain} 
\setcounter{page}{1}
\pagenumbering{arabic}


\input{body}

\input{acknowledgements}


\addcontentsline{toc}{section}{References}
\bibliographystyle{LHCb}
\bibliography{main,standard,LHCb-PAPER,LHCb-DP,LHCb-TDR}

\newpage
\input{Authorship_LHCb-PAPER-2026-015}

\end{document}

%% file: preamble.tex
\usepackage[top=1in, bottom=1.25in, left=1in, right=1in]{geometry}

\usepackage{microtype}
\usepackage{lineno}  
\usepackage{xspace} 
\usepackage{caption} 

\usepackage{graphicx}  
\usepackage{color}
\usepackage{colortbl}
\graphicspath{{./figs/}} 

\usepackage{amsmath} 
\usepackage{amssymb}
\usepackage{amsfonts}
\usepackage{upgreek} 

\newcommand*\patchAmsMathEnvironmentForLineno[1]{%
\expandafter\let\csname old#1\expandafter\endcsname\csname #1\endcsname
\expandafter\let\csname oldend#1\expandafter\endcsname\csname
end#1\endcsname
 \renewenvironment{#1}%
   {\linenomath\csname old#1\endcsname}%
   {\csname oldend#1\endcsname\endlinenomath}%
}
\newcommand*\patchBothAmsMathEnvironmentsForLineno[1]{%
  \patchAmsMathEnvironmentForLineno{#1}%
  \patchAmsMathEnvironmentForLineno{#1*}%
}
\AtBeginDocument{%
\patchBothAmsMathEnvironmentsForLineno{equation}%
\patchBothAmsMathEnvironmentsForLineno{align}%
\patchBothAmsMathEnvironmentsForLineno{flalign}%
\patchBothAmsMathEnvironmentsForLineno{alignat}%
\patchBothAmsMathEnvironmentsForLineno{gather}%
\patchBothAmsMathEnvironmentsForLineno{multline}%
\patchBothAmsMathEnvironmentsForLineno{eqnarray}%
}

\usepackage{hyperxmp}

\usepackage[pdftex,
            pdfauthor={\paperauthors},
            pdftitle={\paperasciititle},
            pdfkeywords={\paperkeywords},
            pdfcopyright={Copyright (C) \papercopyright},
            pdflicenseurl={\paperlicenceurl}]{hyperref}

\usepackage[colorinlistoftodos,textsize=scriptsize]{todonotes}

\usepackage[bottom,flushmargin,hang,multiple]{footmisc}

\usepackage[all]{hypcap} 

\input{lhcb-symbols-def} 

\usepackage{cite} 
\usepackage{mciteplus}

%% file: lhcb-symbols-def.tex
\usepackage{xspace} 
\usepackage{upgreek}

\def\lhcb   {\mbox{LHCb}\xspace}

\def\babar  {\mbox{BaBar}\xspace}

\def\MagUp {\mbox{\em Mag\kern -0.05em Up}\xspace}

\ifthenelse{\boolean{uprightparticles}}%
{

 \def\Pmu         {\ensuremath{\upmu}\xspace}

 \def\Ppi         {\ensuremath{\uppi}\xspace}

 \def\Ppsi        {\ensuremath{\uppsi}\xspace}

 \def\PDelta      {\ensuremath{\Delta}\xspace}                 
 \def\PXi         {\ensuremath{\Xi}\xspace}                 
 \def\PLambda     {\ensuremath{\Lambda}\xspace}                 
 \def\PSigma      {\ensuremath{\Sigma}\xspace}                 
 \def\POmega      {\ensuremath{\Omega}\xspace}                 
 \def\PUpsilon    {\ensuremath{\Upsilon}\xspace}
 \let\oldPi\Pi
 \def\PPi         {\ensuremath{\oldPi}\xspace}

 \def\PB      {\ensuremath{\mathrm{B}}\xspace}                 
                  
 \def\PD      {\ensuremath{\mathrm{D}}\xspace}

 \def\PJ      {\ensuremath{\mathrm{J}}\xspace}                 
 \def\PK      {\ensuremath{\mathrm{K}}\xspace}

 \def\Pb      {\ensuremath{\mathrm{b}}\xspace}                 
 \def\Pc      {\ensuremath{\mathrm{c}}\xspace}                 
 \def\Pd      {\ensuremath{\mathrm{d}}\xspace}

 \def\Pi      {\ensuremath{\mathrm{i}}\xspace}

 \def\Ps      {\ensuremath{\mathrm{s}}\xspace}                 
 \def\Pt      {\ensuremath{\mathrm{t}}\xspace}

 \def\thebaroffset{0.0em}
}
{

 \def\Pmu         {\ensuremath{\mu}\xspace}

 \def\Ppi         {\ensuremath{\pi}\xspace}

 \def\Ppsi        {\ensuremath{\psi}\xspace}                 
                  
 \mathchardef\PDelta="7101
 \mathchardef\PXi="7104
 \mathchardef\PLambda="7103
 \mathchardef\PSigma="7106
 \mathchardef\POmega="710A
 \mathchardef\PUpsilon="7107
 \mathchardef\PPi="7105
                  
 \def\PB      {\ensuremath{B}\xspace}                 
                  
 \def\PD      {\ensuremath{D}\xspace}

 \def\PJ      {\ensuremath{J}\xspace}                 
 \def\PK      {\ensuremath{K}\xspace}

 \def\Pb      {\ensuremath{b}\xspace}                 
 \def\Pc      {\ensuremath{c}\xspace}                 
 \def\Pd      {\ensuremath{d}\xspace}

 \def\Pi      {\ensuremath{i}\xspace}

 \def\Ps      {\ensuremath{s}\xspace}                 
 \def\Pt      {\ensuremath{t}\xspace}

 \def\thebaroffset{0.18em}
}
\newcommand{\offsetoverline}[2][\thebaroffset]{\kern #1\overline{\kern -#1 #2}}%

\makeatletter
\ifcase \@ptsize \relax
  \newcommand{\miniscule}{\@setfontsize\miniscule{4}{5}}
\or
  \newcommand{\miniscule}{\@setfontsize\miniscule{5}{6}}
\or
  \newcommand{\miniscule}{\@setfontsize\miniscule{5}{6}}
\fi
\makeatother

\DeclareRobustCommand{\optbar}[1]{\shortstack{{\miniscule (\rule[.5ex]{1.25em}{.18mm})}
  \\ [-.7ex] $#1$}}

\def\mup        {{\ensuremath{\Pmu^+}}\xspace}
\def\mun        {{\ensuremath{\Pmu^-}}\xspace} 

\def\mumu       {{\ensuremath{\Pmu^+\Pmu^-}}\xspace}

\def\dquark    {{\ensuremath{\Pd}}\xspace}

\def\squark    {{\ensuremath{\Ps}}\xspace}

\def\cquark    {{\ensuremath{\Pc}}\xspace}

\def\bquark    {{\ensuremath{\Pb}}\xspace}

\def\tquark    {{\ensuremath{\Pt}}\xspace}

\def\pion   {{\ensuremath{\Ppi}}\xspace}

\def\pip    {{\ensuremath{\pion^+}}\xspace}
\def\pim    {{\ensuremath{\pion^-}}\xspace}

\def\kaon    {{\ensuremath{\PK}}\xspace}

\def\KorKbar {\kern \thebaroffset\optbar{\kern -\thebaroffset \PK}{}\xspace}

\def\Kp      {{\ensuremath{\kaon^+}}\xspace}
\def\Km      {{\ensuremath{\kaon^-}}\xspace}

\def\D       {{\ensuremath{\PD}}\xspace}

\def\DorDbar {\kern \thebaroffset\optbar{\kern -\thebaroffset \PD}\xspace}
\def\Dz      {{\ensuremath{\D^0}}\xspace}

\def\Dp      {{\ensuremath{\D^+}}\xspace}
\def\Dm      {{\ensuremath{\D^-}}\xspace}

\def\DpDm    {\ensuremath{\Dp {\kern -0.16em \Dm}}\xspace}

\def\Dstarp  {{\ensuremath{\D^{*+}}}\xspace}

\def\B       {{\ensuremath{\PB}}\xspace}

\def\BorBbar {\kern \thebaroffset\optbar{\kern -\thebaroffset \PB}\xspace}

\def\Bd      {{\ensuremath{\B^0}}\xspace}

\def\BdorBdbar {\kern \thebaroffset\optbar{\kern -\thebaroffset \Bd}\xspace}
\def\Bu      {{\ensuremath{\B^+}}\xspace}
\def\Bub     {{\ensuremath{\B^-}}\xspace}
\def\Bp      {{\ensuremath{\Bu}}\xspace}
\def\Bm      {{\ensuremath{\Bub}}\xspace}

\def\Bs      {{\ensuremath{\B^0_\squark}}\xspace}

\def\BsorBsbar {\kern \thebaroffset\optbar{\kern -\thebaroffset \Bs}\xspace}
\def\Bc      {{\ensuremath{\B_\cquark^+}}\xspace}

\def\jpsi     {{\ensuremath{{\PJ\mskip -3mu/\mskip -2mu\Ppsi}}}\xspace}

\def\Y#1S{\ensuremath{\PUpsilon{(#1S)}}\xspace}

\def\LorLbar     {\kern \thebaroffset\optbar{\kern -\thebaroffset \PLambda}\xspace}

\newcommand{\decay}[2]{\ensuremath{\mathinner{#1\!\to #2}}\xspace}

\def\to                 {\ensuremath{\rightarrow}\xspace}

\def\qsq       {{\ensuremath{q^2}}\xspace}

\def\CP                {{\ensuremath{C\!P}}\xspace}

\def\Vtd  {{\ensuremath{V_{\tquark\dquark}^{\phantom{\ast}}}}\xspace}

\def\AFB      {\ensuremath{A_{\mathrm{FB}}}\xspace}

\def\AT#1     {\ensuremath{A_{\mathrm{T}}^{#1}}\xspace}           

\def\C#1      {\ensuremath{\mathcal{C}_{#1}}\xspace}                       
\def\Cp#1     {\ensuremath{\mathcal{C}_{#1}^{'}}\xspace}                    
\def\Ceff#1   {\ensuremath{\mathcal{C}_{#1}^{\mathrm{(eff)}}}\xspace}        
\def\Cpeff#1  {\ensuremath{\mathcal{C}_{#1}^{'\mathrm{(eff)}}}\xspace}       
\def\Ope#1    {\ensuremath{\mathcal{O}_{#1}}\xspace}                       
\def\Opep#1   {\ensuremath{\mathcal{O}_{#1}^{'}}\xspace}                    

\newcommand{\nospaceunit}[1]{\ensuremath{\text{#1}}}       
\newcommand{\aunit}[1]{\ensuremath{\text{\,#1}}}       

\newcommand{\tev}{\aunit{Te\kern -0.1em V}\xspace}
\newcommand{\gev}{\aunit{Ge\kern -0.1em V}\xspace}
\newcommand{\mev}{\aunit{Me\kern -0.1em V}\xspace}
\newcommand{\kev}{\aunit{ke\kern -0.1em V}\xspace}
\newcommand{\ev}{\aunit{e\kern -0.1em V}\xspace}
 
\newcommand{\mevc}{\ensuremath{\aunit{Me\kern -0.1em V\!/}c}\xspace}
\newcommand{\gevc}{\ensuremath{\aunit{Ge\kern -0.1em V\!/}c}\xspace}
\newcommand{\mevcc}{\ensuremath{\aunit{Me\kern -0.1em V\!/}c^2}\xspace}
\newcommand{\gevcc}{\ensuremath{\aunit{Ge\kern -0.1em V\!/}c^2}\xspace}
\newcommand{\gevgevcccc}{\ensuremath{\gev^2\!/c^4}\xspace} 

\def\mum  {\ensuremath{\,\upmu\nospaceunit{m}}\xspace}

\def\fb   {\ensuremath{\aunit{fb}}\xspace}
\def\invfb   {\ensuremath{\fb^{-1}}\xspace}

\newcommand{\chisq}{\ensuremath{\chi^2}\xspace}

\newcommand{\chisqip}{\ensuremath{\chi^2_{\text{IP}}}\xspace}

\def\deriv {\ensuremath{\mathrm{d}}}

\def\gsim{{~\raise.15em\hbox{$>$}\kern-.85em
          \lower.35em\hbox{$\sim$}~}\xspace}
\def\lsim{{~\raise.15em\hbox{$<$}\kern-.85em
          \lower.35em\hbox{$\sim$}~}\xspace}

\def\pt         {\ensuremath{p_{\mathrm{T}}}\xspace}

\def\ptot       {\ensuremath{p}\xspace}

\def\mrad{\aunit{mrad}\xspace}

\def\evtgen     {\mbox{\textsc{EvtGen}}\xspace}

\def\geant      {\mbox{\textsc{Geant4}}\xspace}

\def\photos     {\mbox{\textsc{Photos}}\xspace}

\def\pythia     {\mbox{\textsc{Pythia}}\xspace}

\def\tell1  {TELL1\xspace}
\def\ukl1   {UKL1\xspace}

\newcommand{\ie}{\mbox{\itshape i.e.}\xspace}

\newcommand{\lhcborcid}[1]{\href{https://orcid.org/#1}{\hspace*{0.1em}\raisebox{-0.45ex}{\includegraphics[width=1em]{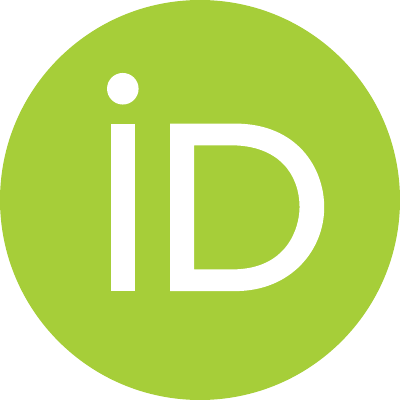}}}}


%% file: title-LHCb-PAPER.tex

\begin{titlepage}
\pagenumbering{roman}

\vspace*{-1.5cm}
\centerline{\large EUROPEAN ORGANIZATION FOR NUCLEAR RESEARCH (CERN)}
\vspace*{1.5cm}
\noindent
\begin{tabular*}{\linewidth}{lc@{\extracolsep{\fill}}r@{\extracolsep{0pt}}}
\ifthenelse{\boolean{pdflatex}}
{\vspace*{-1.5cm}\mbox{\!\!\!\includegraphics[width=.14\textwidth]{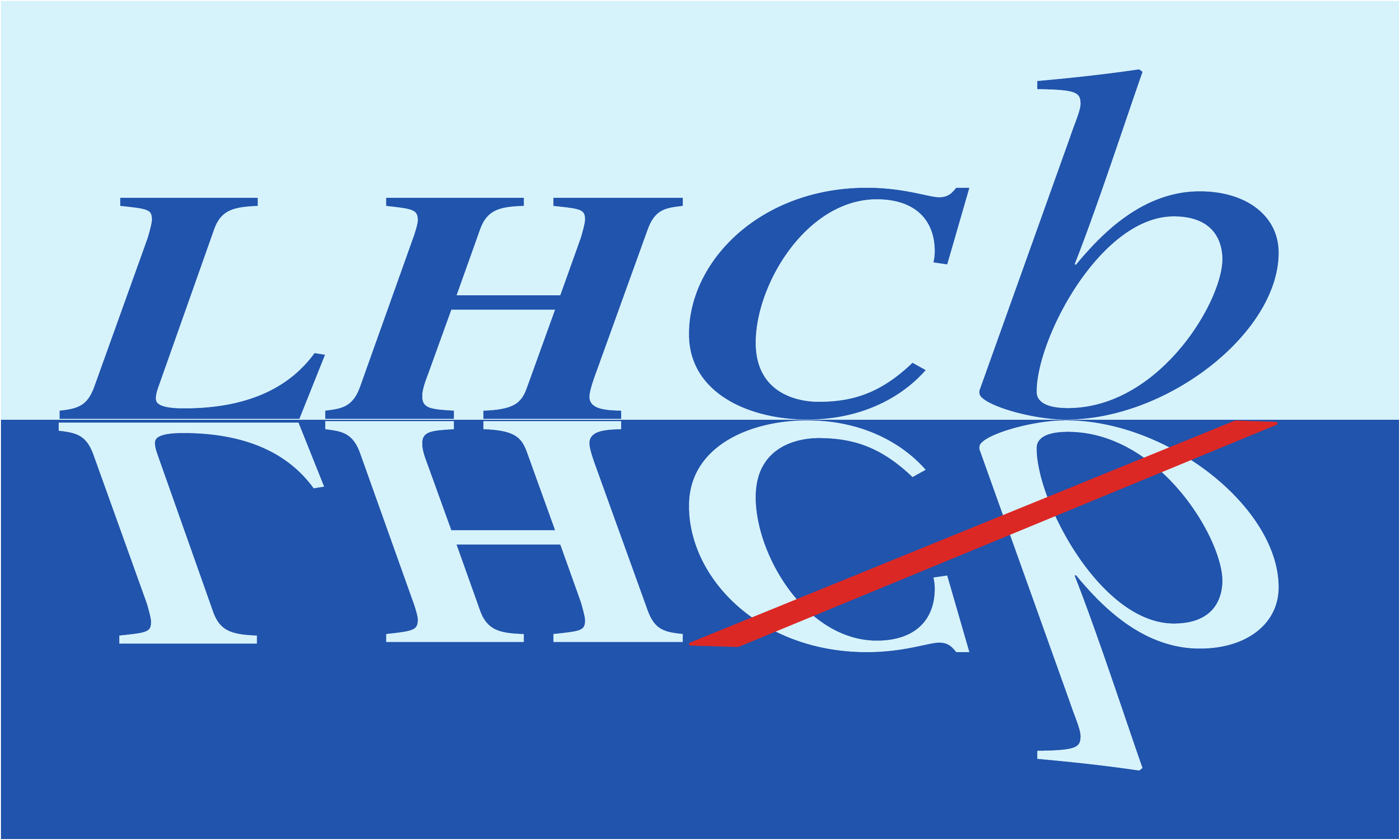}} & &}%
{\vspace*{-1.2cm}\mbox{\!\!\!\includegraphics[width=.12\textwidth]{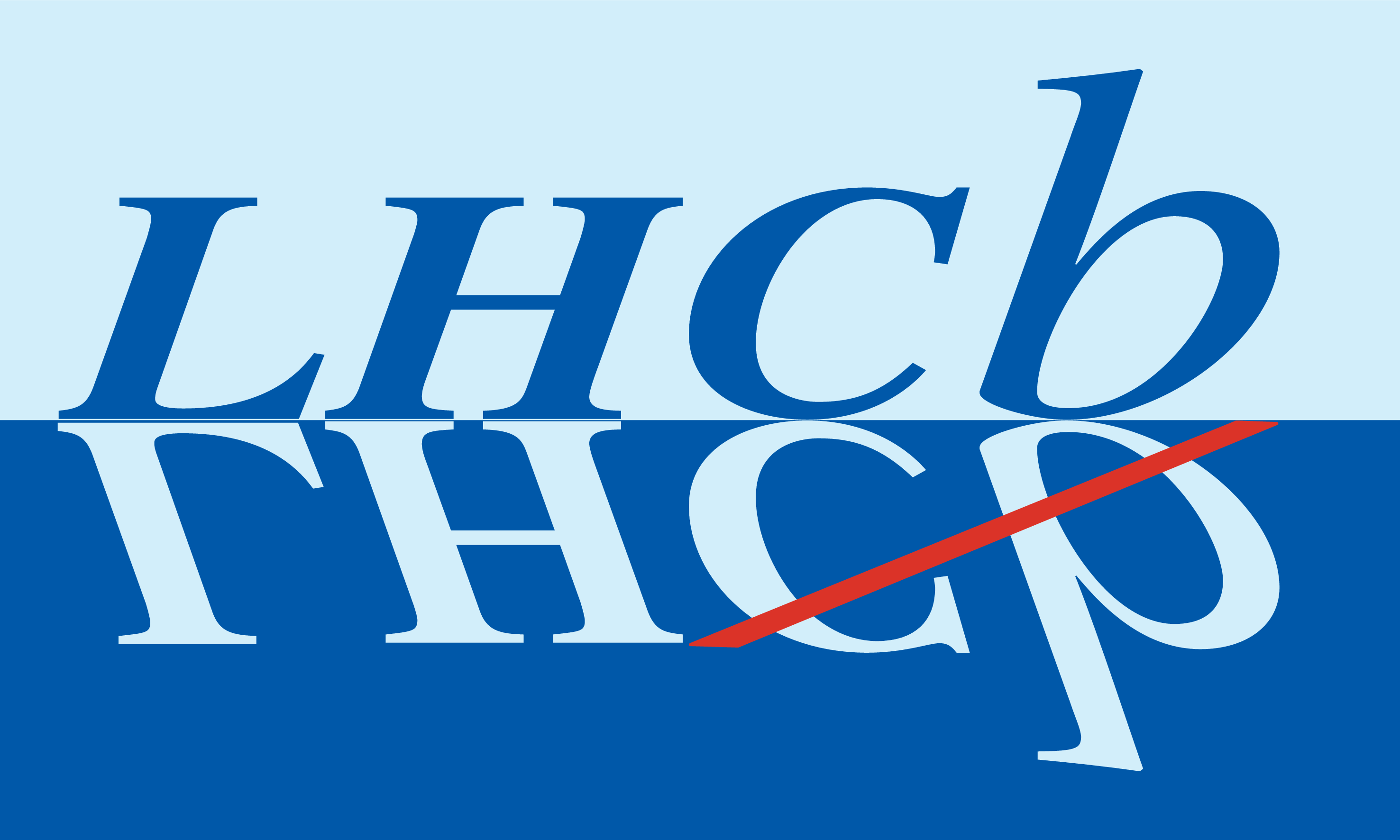}} & &}%
\\
 & & CERN-EP-2026-105 \\  
 & & LHCb-PAPER-2026-015 \\  
 & & August 03, 2026 \\
 & & \\
\end{tabular*}

\vspace*{4.0cm}

{\normalfont\bfseries\boldmath\huge
\begin{center}
  \papertitle 
\end{center}
}

\vspace*{2.0cm}

\begin{center}
\paperauthors\footnote{Authors are listed at the end of this paper.}
\end{center}

\vspace{\fill}

\begin{abstract}
  \noindent
This paper presents the first measurement of the forward-backward asymmetry, $A_{\rm FB}$, and the flat term, $F_{H}$, that parameterise the angular distribution of the $B^+\to\pi^+\mu^+\mu^-$ decay. 
The proton-proton collision dataset used in the analysis corresponds to an integrated luminosity of 9\,fb$^{-1}$, collected with the LHCb experiment between 2011 and 2018. 
The analysis is performed in two intervals of dimuon mass squared, one above and one below the region containing the $J\mskip -3mu/\mskip -2mu\psi$ and $\psi(2S)$ narrow charmonium resonances.
The Standard Model predictions lie within the obtained 68\% confidence level interval in the high-mass and within the 99\% interval in the low-mass region.
\end{abstract}

\vspace*{2.0cm}

\begin{center}
Published in \href{https://doi.org/10.1007/JHEP08(2026)004}{JHEP 08 (2026) 004}
\end{center}

\vspace{\fill}

{\footnotesize 
\centerline{\copyright~\papercopyright. \href{\paperlicenceurl}{\paperlicence}.}}
\vspace*{2mm}

\end{titlepage}


\newpage
\setcounter{page}{2}
\mbox{~}

%% file: body.tex
\section{Introduction}
\label{sec:Introduction}

\begin{figure}[b]
    \centering
    \includegraphics[width=\textwidth]{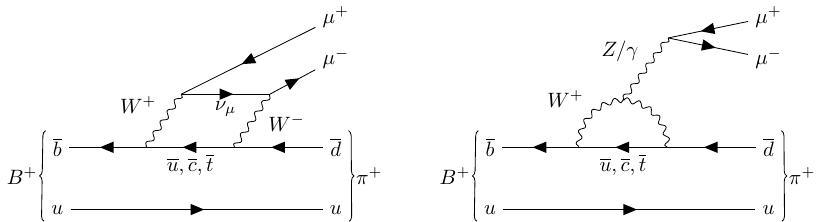}
    \caption{Leading-order Feynman diagrams for the \BuPiMuMu decay in the SM.
    }
    \label{fig:feynman}
\end{figure}

In the Standard Model of particle physics (SM), the \BuPiMuMu decay is mediated by a \bquark- to \dquark-quark flavour-changing neutral-current transition,\footnote{
The inclusion of charge-conjugate processes is implied throughout this paper unless explicitly stated.
} 
with the leading-order Feynman diagrams shown in Fig.~\ref{fig:feynman}.
The dominant contribution arises from diagrams with an internal \tquark-quark loop, since the large \tquark mass enhances its virtual effects.
The small magnitude of the \Vtd element of the Cabibbo--Kobayashi--Maskawa (CKM) quark-mixing matrix~\cite{Cabibbo:1963yz,Kobayashi:1973fv} suppresses the branching fraction of the decay to $\mathcal{O}\left(10^{-8}\right)$. 
The suppressed SM branching fraction makes the \BuPiMuMu decay sensitive to contributions from particles arising in extensions of the SM. 
Non-SM contributions could alter observables such as the rate and angular distribution of the decay or differences between \Bp and \Bm decays. 
While previous measurements of the branching fraction and \CP asymmetry~\cite{LHCb-PAPER-2012-020,LHCb-PAPER-2015-035} are consistent with SM predictions~\cite{Ali:2013zfa,Hambrock:2015wka,FermilabLattice:2015cdh,Khodjamirian:2017fxg,Biswas:2022lhu,Rusov:2019ixr,Bause:2022rrs}, the angular distribution of the \BuPiMuMu decay has not previously been measured. 

\begin{figure}
    \centering
    \includegraphics[width=0.9\textwidth]{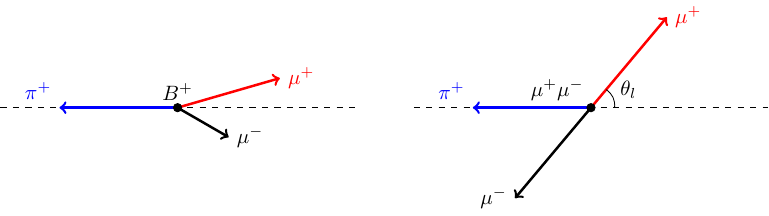}
    \caption{
    Illustration of the \BuPiMuMu decay in the rest frame of (left)~the \Bp~meson and (right)~the dimuon system. The angle $\Thetal$ is defined as the angle between the direction of the \mup and the direction opposite the pion in the dimuon rest frame.
    }
    \label{fig:angle}
\end{figure}

The differential decay rate of the \BuPiMuMu process can be parameterised in terms of the dimuon mass squared, \qsq, and the angle, \Thetal, defined as the angle between the \mup direction and the direction opposite to the pion in the dimuon rest frame (see Fig.~\ref{fig:angle}). 
Normalising by the total decay rate $\Gamma$ gives~\cite{Bobeth:2007dw, Ali:1999mm}
\begin{align}
\frac{1}{\Gamma} \frac{\deriv\Gamma}{\deriv\CosThetal} = \frac{3}{4}(1 - \FH)(1 - \cos^{2}\Thetal) + \frac{\FH}{2} + \AFB \CosThetal\,.
\label{eq:angular}
\end{align} 
This expression depends on two parameters, the forward-backward asymmetry of the dimuon system, \AFB, and \FH, which characterises the flatness of the \CosThetal distribution. 
These parameters are sensitive to scalar, pseudoscalar and tensor contributions to the decay width~\cite{Beaujean:2015gba}. 
In the SM, \FH is small in the regions away from the endpoints of the \qsq\ spectrum and $\AFB = 0$.  
For the probability density of the angular distribution to remain positive definite, $|\AFB| \leq \FH/2$ is required. 
Small deviations from Eq.~\eqref{eq:angular} due to quantum electrodynamics effects~\cite{Isidori:2020acz} can be neglected given the size of currently available data samples. 

The angular distribution of the \BuPiMuMu decay is studied for the first time in this paper using a dataset of proton-proton~($pp$) collisions, corresponding to a total integrated luminosity of 9\invfb, collected with the LHCb experiment at centre-of-mass energies of 7, 8 and 13\tev.  
The data are split into two regions of \qsq, covering $1.1< \qsq <6.0$\gevgevcccc and $15.0 < \qsq < 22.0$\gevgevcccc. 
While no measurements have previously been made of the angular distribution of the \BuPiMuMu decay, the \babar, Belle, CDF, CMS and LHCb collaborations have previously studied the corresponding distribution for the \BuKMuMu decay~\cite{BaBar:2006tnv,Belle:2009zue,CDF:2011tds,CMS:2018qih,LHCb-PAPER-2014-007}.
This decay has a much larger branching fraction due to the relative magnitude of the $V_{ts}$ element of the CKM matrix, compared to $V_{td}$, and in the SM is expected to have a similar angular distribution. 
The results of the \BuKMuMu analyses are consistent with the SM, where \AFB and \FH are consistent with zero. 
These measurements have been used to set the best available limits on tensor contributions to $b\to s$ processes~\cite{Beaujean:2015gba}.

\section{Detector and simulation}
\label{sec:Detector} 

The \lhcb detector~\cite{LHCb-DP-2008-001,LHCb-DP-2014-002} is a single-arm forward spectrometer covering the \mbox{pseudorapidity} range $2 < \eta < 5$,
designed for the study of particles containing \bquark or \cquark quarks. 
The detector used to collect the data analysed in this paper includes a high-precision tracking system consisting of a silicon-strip vertex detector surrounding the $pp$ interaction region~\cite{LHCb-DP-2014-001}, a large-area silicon-strip detector located upstream of a dipole magnet with a bending power of about $4\,{\mathrm{T\,m}}$, and three stations of silicon-strip detectors and straw drift tubes~\cite{LHCb-DP-2013-003,LHCb-DP-2017-001} placed downstream of the magnet.
The tracking system provides a measurement of the momentum, \ptot, of charged particles with a relative uncertainty that varies from 0.5\% at low momentum to 1.0\% at 200\gevc.
The minimum distance of a track to a primary $pp$ collision vertex (PV), the impact parameter (IP), is measured with a resolution of $(15+29/\pt)\mum$, where \pt is the component of the momentum transverse to the beam, in\,\gevc.
Different types of charged hadrons are distinguished using information from two ring-imaging Cherenkov detectors~\cite{LHCb-DP-2012-003}. 
Photons, electrons and hadrons are identified by a calorimeter system consisting of scintillating-pad and preshower detectors, an electromagnetic and a hadronic calorimeter~\cite{LHCb-DP-2020-001}. 
Muons are identified by a system composed of alternating layers of iron and multiwire proportional chambers~\cite{LHCb-DP-2012-002}. 
The online event selection is performed by a trigger~\cite{LHCb-DP-2012-004}, which consists of a hardware stage, based on information from the calorimeter and muon systems, followed by a software stage, which applies a full event reconstruction. 
Triggered data further undergo a centralised, offline processing step to deliver physics-analysis-ready data across the entire \lhcb physics programme~\cite{Stripping}.

Simulated events are used to understand the impact of the detector and subsequent selection requirements on the angular distribution of the signal \BuPiMuMu decays. 
They are also used to model the contribution from a variety of backgrounds, such as from the \BuKMuMu decay, and to optimise the signal selection process. 
In the simulation, $pp$ collisions are generated using \pythia~\cite{Sjostrand:2007gs} with a specific \lhcb configuration~\cite{LHCb-PROC-2010-056}. 
Decays of unstable particles are described by \evtgen~\cite{Lange:2001uf}, in which final-state radiation is generated using \photos~\cite{davidson2015photos}.
The angular and \qsq\ distributions of \BuPiMuMu and \BuKMuMu decays are generated according to the SM, following the description in Ref.~\cite{Ali:1999mm} with form factors from Ref.~\cite{Ball:2004ye}.
The interaction of the generated particles with the detector, and its response, are implemented using the \geant toolkit~\cite{Allison:2006ve, *Agostinelli:2002hh} as described in Ref.~\cite{LHCb-PROC-2011-006}. 
The simulated samples are corrected for known data-simulation differences in track reconstruction, trigger and muon-identification efficiencies using a tag-and-probe approach with \JpsiMuMu decays~\cite{LHCb-DP-2013-002,LHCb-PUB-2014-039,LHCb-PUB-2016-021}. 
Kaon- and pion-identification efficiencies are corrected using $\decay{\Dstarp}{\Dz(\Km\pip)\pip}$ decays~\cite{LHCb-PUB-2016-021}, which can be selected in data without particle-identification requirements. 
Simulated samples are then weighted such that the production kinematics of the \Bp mesons in the simulation are consistent with the observed momentum distributions of \Bp mesons reconstructed through \BuJpsiK decays in data. 

\section{Candidate selection} 
\label{sec:Selection}

Events are required by the hardware trigger to contain at least one muon with high \pt, with the minimum threshold varying in the range 1200--2200\mevc depending on the data-taking period. 
In the software trigger, at least one of the final-state particles is required to have $\pt > 800\mevc$ and an IP greater than $100\mum$ with respect to all PVs in the event. 
A multivariate algorithm~\cite{BBDT,LHCb-PROC-2015-018} 
is then used to identify events containing secondary vertices formed by two or three tracks that are consistent with the decay of a \bquark~hadron.

In the offline selection, which largely follows procedures developed in Ref.~\cite{LHCb-PAPER-2023-004}, \Bp~candidates are formed from the combination of two oppositely charged tracks that are identified as muons and a third track identified as a pion. 
Both muons are required to have $\pt > 300\mevc$. 
Each track is required to have a good track-fit quality, to be within the acceptance of the particle-identification systems, and have a significant IP with respect to all PVs in the event. 
The three tracks are required to be consistent with originating from a single point that is displaced from every PV.  
The \Bp candidate is required to have a \PiMuMu mass in the range $5180 < m(\PiMuMu) < 5780$\mevcc and to be consistent with originating from its associated PV, \ie\ the PV with the smallest \chisqip, defined as the difference in the vertex-fit \chisq of the PV reconstructed with and without the candidate.

Combinatorial background, comprising tracks originating from two or more different particle decays, is suppressed using a boosted decision tree classifier~\cite{Breiman,Chen:2016:XST:2939672.2939785}.
The classifier is trained using samples of simulated decays for signal and background candidates from sidebands of the data with \PiMuMu mass in the range \mbox{$5550 < m(\PiMuMu) < 6000\mevcc$} or \mbox{$6500 < m(\PiMuMu) < 7000$\mevcc}. 
The range \mbox{$6000 < m(\PiMuMu) < 6500$\mevcc} is excluded to avoid potential contributions from resonant or nonresonant \BcPiMuMu decays~\cite{LHCb-PAPER-2023-037,LHCb-PAPER-2024-026}. 
The classifier uses a variety of topological and kinematic variables that are characteristic of \Bp meson decays. 
These include the muon and pion momenta, the muon \chisqip, the \Bp candidate vertex-fit \chisq and the angle between the \Bp momentum-vector and the vector connecting
the \Bp decay vertex and its associated PV. 
The classifier working point and particle-identification requirements applied to the pion are optimised simultaneously to maximise the expected significance of the \BuPiMuMu signal. 

The largest potential source of background from a specific decay arises from the \BuKMuMu channel, where the kaon is incorrectly identified as a pion.  
Before applying particle-identification requirements, this background is 25 times more abundant than the signal~\cite{PDG2024}. 
Residual background contributions arising from \BuPiPiPi\ and \BuKPiPi decays amount to approximately 1\% of the signal yield. 
These contributions are neglected in the analysis and accounted for as a source of systematic uncertainty.
Backgrounds are also present from \BuJpsiK\ and \BuJpsiPi decays, where the kaon or pion is incorrectly identified as a muon, while the positive muon from the \JpsiMuMu decay is misidentified as a pion. 
These are removed by vetoing candidates with a $\pip\mun$ mass, where the muon mass hypothesis is assigned to the pion, within 20\mevcc of the known \jpsi mass~\cite{PDG2024}; this is subsequently referred to as the \jpsi veto.   
Decays of \Bd, \Bu and \Bs mesons to final states containing three tracks and additional unreconstructed particles predominantly populate the \PiMuMu mass region below 5180\mevcc.
These contributions are removed by the requirement on the \PiMuMu mass and are therefore not considered further.
To suppress the most significant source of semileptonic background with missing neutrinos, candidates are vetoed if they have a $\pip\mun$ mass consistent with a \D meson. 
The largest residual contribution from decays with unreconstructed particles within the considered \PiMuMu mass range is from \BsfMuMu, \decay{f_0(980)}{\pip\pim} decays. 
Given its small branching fraction~\cite{LHCb-PAPER-2014-063}, this background is neglected in the analysis and accounted for as a source of systematic uncertainty. 

Samples of \BuJpsiPi and \BuJpsiK with \JpsiMuMu decays are retained as control channels. 
The \BuJpsiPi decay is selected using the same criteria as the \BuPiMuMu signal, except for the \qsq selection that is replaced by requiring the $\mup\mun$ mass to be within 100\mevcc of the known \jpsi mass. 
The \BuJpsiK decay is selected using an alternative set of particle-identification requirements, designed to select kaons and reject pions.

\section{Mass and angular model} 
\label{sec:Model}

In each \qsq\ interval, the observables \AFB and \FH are determined by performing a simultaneous unbinned maximum-likelihood fit to the \PiMuMu mass and \CosThetal distributions of the selected candidates. 
The signal angular distribution is modelled by Eq.~\eqref{eq:angular}, multiplied by a function describing the angular efficiency of the reconstruction and selection process. 
The angular resolution of the detector is better than 1\mrad and can therefore be neglected, but is considered as a source of systematic uncertainty. 

The \PiMuMu mass and \CosThetal distributions are assumed to be statistically independent in the fits. 
For the signal, this assumption is validated using simulation and large data samples of \BuJpsiPi and \BuJpsiK decays. 
For the combinatorial background, the assumption is validated using sideband data. 
For the \BuKMuMu background, a dependence between the two observables arises from the wrong hadron mass assignment. 
Nevertheless, it is also assumed that the mass and angular distributions of the \BuKMuMu component factorise in the fit, and an associated systematic uncertainty is assigned. 

\begin{figure}[tb]
    \centering
    \includegraphics[width=0.49\linewidth]{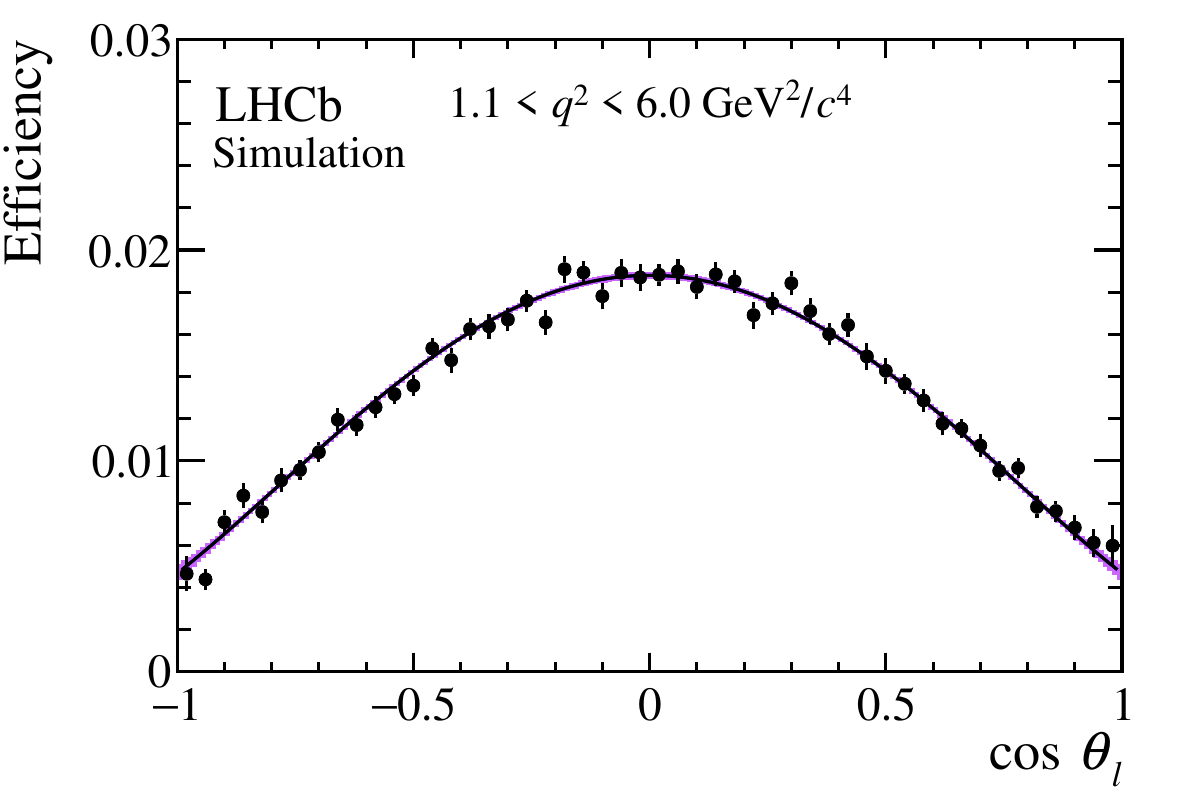}
    \includegraphics[width=0.49\linewidth]{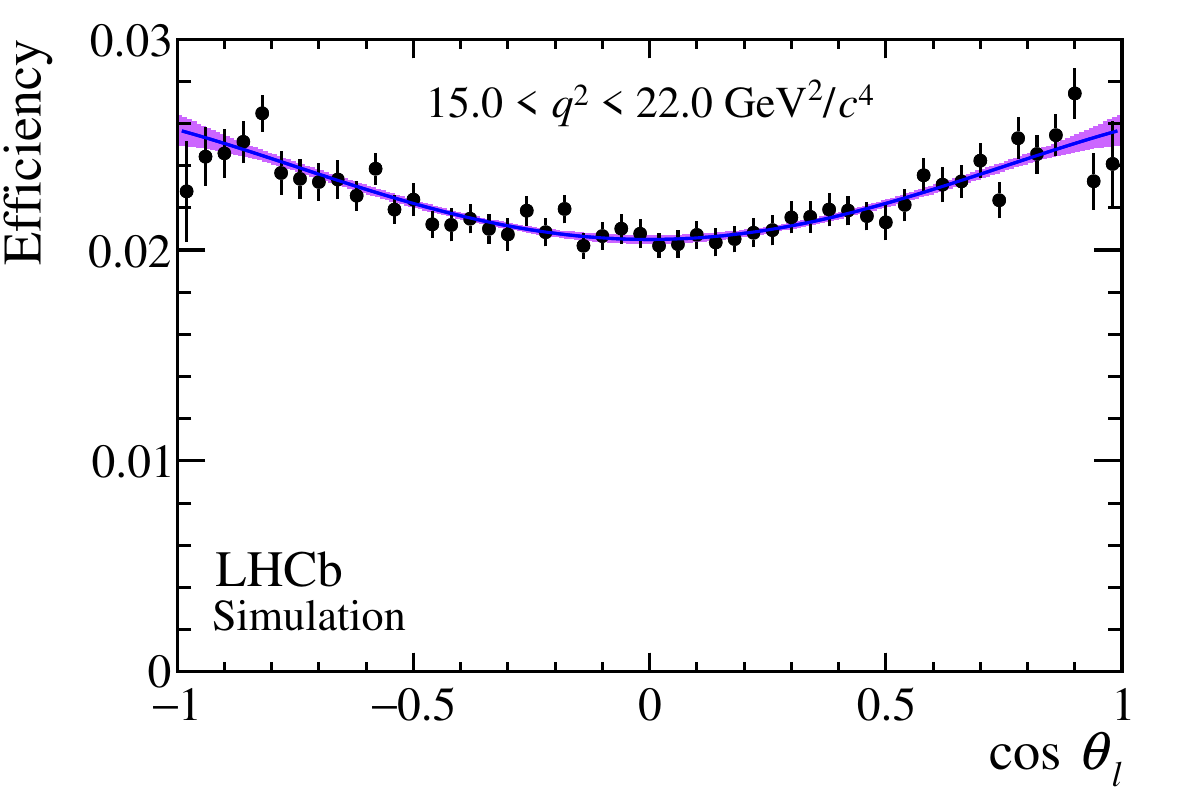}
    \caption{
    Efficiency as a function of \CosThetal for \BuPiMuMu decays in simulation in the \qsq\ regions (left)~$1.1 < \qsq < 6.0$\gevgevcccc and (right)~$15.0 < \qsq < 22.0$\gevgevcccc. 
    The projections of fits using a fourth-order polynomial parameterisation, containing only even terms, are superimposed, with the uncertainty on the obtained efficiency function shown as a band.
    }
    \label{fig:pimumu:efficiency:syst:fit}
\end{figure}

Figure~\ref{fig:pimumu:efficiency:syst:fit} shows the efficiency of the reconstruction and selection processes, without the \jpsi veto applied, as a function of \CosThetal in the two \qsq\ ranges on simulated \BuPiMuMu decays. 
The efficiency is well described in each \qsq\ interval by a fourth-order Legendre polynomial that only includes even terms and is therefore symmetric around $\CosThetal = 0$. 
The variation of \BuPiMuMu kinematics causes the selection efficiency to differ between the low- and high-\qsq intervals. 
The main sources of inefficiency are related to the requirement that the muons need sufficient momentum to reach and traverse the muon system and to satisfy the hardware trigger. 
At low \qsq, the track and muon reconstruction requirements disfavour decays where $\Thetal$ is close to 0 or $\pi$, where one muon is soft in the \Bp rest frame. 
At intermediate and high \qsq, the hardware trigger requirements on the \pt of the muon candidates disfavour candidates with $\Thetal\sim\pi/2$, which typically have a smaller maximum muon momentum.

The simulation is generated with an SM \qsq\ distribution~\cite{Ali:1999mm, Ball:2004ye}, and potential variations of the efficiency profile within each \qsq\ interval are considered as a source of systematic uncertainty.
The \jpsi veto on the $\pip\mun$ mass removes a narrow (\qsq dependent) region of \CosThetal, which cannot be modelled by a polynomial. 
The efficiency is multiplied by a discontinuous function to account for this effect.

The signal mass distribution is described by the sum of a Gaussian function with power-law tails on both sides of the distribution~\cite{Skwarnicki:1986xj} and a second Gaussian function, with common peak positions. 
The tail parameters and relative fractions of the two functions are fixed from fits to simulation performed in each $q^2$ interval separately. 
In the fits to data, an offset in the position of the peak and a scale factor on the width of the distribution, due to momentum-scale uncertainties, are included and determined from fits to the $\Kp\mumu$ mass distribution of selected \BuJpsiK decays. 

The background from misidentified \BuKMuMu decays is modelled using simulation, as the distribution in data is known to be consistent with SM expectations~\cite{LHCb-PAPER-2014-007}. 
The angular distribution is described by a fourth-order Chebyshev polynomial function, with no contribution from odd terms.  
The mass distribution of the misidentified background is described by a Gaussian function with power-law tails on both sides of the peak.
The shape of this background is fixed from simulation in each $q^2$ interval separately, aside from factors accounting for data-simulation differences in the width and peak-position that are shared with the signal. 

Combinatorial background is described by an exponential function in \PiMuMu mass and a second-order Chebyshev polynomial function in \CosThetal. 
The parameters of these distributions are free to vary in the fit to data. 

Owing to the physical boundary $|\AFB| \leq \FH/2$ and the small size of the samples in the two \qsq\ intervals, the Feldman--Cousins approach~\cite{Feldman:1997qc} is used to estimate confidence intervals on \AFB and \FH. 
At each fixed value of \FH and \AFB, the best-fit value for the nuisance parameters is used when generating pseudoexperiments. 
When presenting intervals on \AFB (\FH), \FH (\AFB) is treated as a nuisance parameter. 

A fit is performed to the \BuJpsiPi control sample in order to validate the analysis strategy. 
The model used to describe \BuJpsiPi decays is identical to that of the signal but with the vetoes removed and the efficiency parameterised over the appropriate \qsq\ range. 
The \BuJpsiPi decay has a known angular distribution, proportional to $\sin^2\Thetal$. This is a signal-like distribution but with $\AFB=0$ and $\FH=0$. 
A fit is also performed to the sample of \BuJpsiK decays. In this fit only contributions from signal and combinatorial background are considered. 

\section{Systematic uncertainties} 
\label{sec:Systematics} 

The impact of various sources of systematic uncertainty is estimated using pseudoexperiments generated with sample sizes three orders of magnitude larger than the data.  
In most cases, the experiments are generated according to an alternative hypothesis and then fit using the baseline analysis strategy. 
A systematic uncertainty is conservatively assigned based on the bias on the fitted observables added together with the uncertainty on the observable in quadrature. 
Exceptions to this process are discussed below. 
The different sources of systematic uncertainty are summarised in Table~\ref{tab:systematics}. 

\begin{table}[!tb] 
\caption{
Sources of systematic uncertainty on \FH and \AFB. 
The different contributions are described in the text, and the total uncertainty is obtained by adding the individual sources of uncertainty in quadrature. 
}
\centering
\begin{tabular}{r|cc | cc}
\toprule
Range $[\!\gevgevcccc]$
& \multicolumn{2}{c|}{$1.1 < \qsq < 6.0$} & \multicolumn{2}{c}{$15.0 < \qsq < 22.0$}\\
Source $\backslash$ Observable & \FH & \AFB & \FH & \AFB\\
\midrule
Simulation \qsq dependence      & 0.12 & 0.01 & 0.01 & 0.01 \\
Angular efficiency order        & 0.01 & 0.01 & 0.01 & 0.01 \\
Simulation sample size          & 0.01 & 0.01 & 0.01 & 0.02 \\
Data-simulation corrections     & 0.01 & 0.03 & 0.02 & 0.04 \\
Truth-matching criteria         & 0.01 & 0.01 & 0.01 & 0.02 \\
Angular resolution              & 0.01 & 0.01 & 0.01 & 0.02 \\
\BuKMuMu background model       & 0.01 & 0.01 & 0.01 & 0.02 \\
Combinatorial background model  & 0.01 & 0.01 & 0.01 & 0.02 \\
Hadronic backgrounds            & 0.04 & 0.01 & 0.02 & 0.01 \\
\BsfMuMu background             & 0.01 & 0.01 & 0.02 & 0.02 \\
\midrule                                              
Total                           & 0.13 & 0.04 & 0.04 & 0.07 \\
\bottomrule
\end{tabular}
\label{tab:systematics} 
\end{table} 

Five types of systematic uncertainty are considered in connection with the signal efficiency model. 
The uncertainty associated with the knowledge of the \qsq\ spectrum of the signal is the largest source of systematic uncertainty on \FH. 
To assign this uncertainty, the simulation is weighted to match the \qsq\ spectrum observed in data~\cite{LHCb-PAPER-2026-016} and a pseudoexperiment is generated according to the angular efficiency of the weighted simulation.
The uncertainty accounting for the efficiency parameterisation is assigned by considering alternative models: one that allows odd terms and another that allows for an additional term of order six. 
The uncertainty due to the limited size of the simulated samples used to derive the efficiency model is evaluated by bootstrapping~\cite{efron:1979} the simulation and deriving a set of varied models. 
Pseudoexperiments are then generated from each efficiency model and fit with the baseline model. 
The spread in values of the observables from this set of pseudoexperiments is assigned as a systematic uncertainty. 
A similar approach is used to derive an uncertainty associated with the data-simulation corrections. 
Potential imperfections in the matching of reconstructed candidates to true decays in the simulation are a further source of uncertainty. 
To assess this uncertainty, a pseudoexperiment is generated according to an efficiency model derived with a more relaxed set of matching criteria on the pion and muon tracks. 

Five sources of systematic uncertainty are considered related to choices made in modelling the mass and angular distributions of the signal and background components. 
To assess the impact of neglecting the signal angular resolution, a pseudoexperiment is generated with the \Thetal distribution broadened according to the resolution obtained in the simulation. 
The systematic uncertainty associated to the assumption that the \PiMuMu mass and \CosThetal distributions factorise for misidentified \BuKMuMu decays is assessed by producing a pseudoexperiment in which the \BuKMuMu background component is modelled nonparametrically with events sampled from simulation. 
The uncertainty due to the combinatorial background model is assessed by relaxing the classifier requirement and fitting the upper mass sideband of the data using a fourth-order polynomial function. 
The shape parameters from this fit are subsequently used when generating the combinatorial background component of a pseudoexperiment. 
Finally, uncertainties due to neglecting hadronic backgrounds from \BuKPiPi, \BuPiPiPi and \BsfMuMu decays are determined by including additional background components in the pseudoexperiment generation. 
The mass and angular distributions of the hadronic background components are based on simulated \BuKPiPi and \BuPiPiPi decays that are weighted to reproduce the Dalitz structure of the decays measured by BaBar~\cite{BaBar:2008lpx} and LHCb~\cite{LHCb-PAPER-2019-017}, respectively.
The yield of the hadronic background is determined using control regions of the data with relaxed muon-identification requirements. 

\section{Results} 
\label{sec:Results}

\begin{figure}[!tb]
\centering
\includegraphics[width=\linewidth]{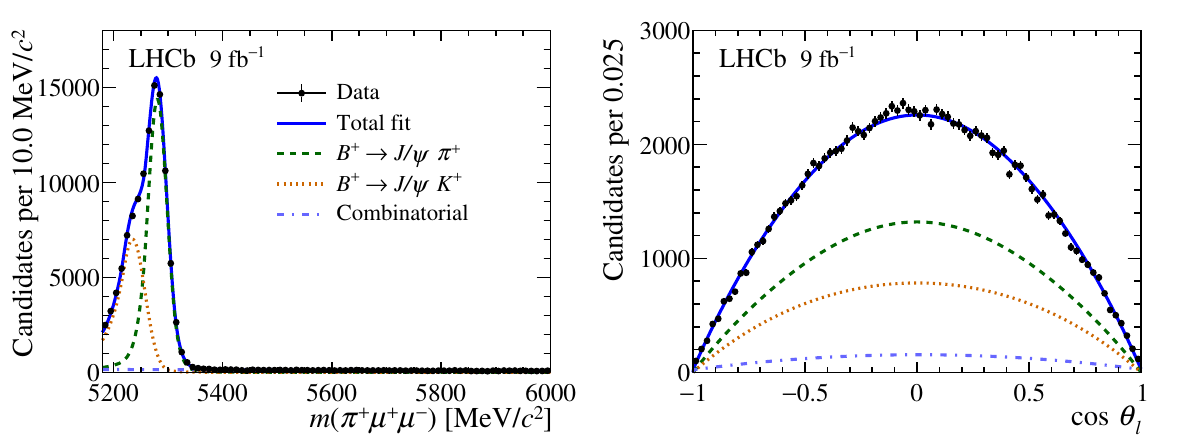}
\caption{ 
Distribution of (left)~the \PiMuMu mass and (right)~\CosThetal for selected \BuJpsiPi candidates. 
The data are superimposed on the result of the fit described in the text. 
}
\label{fig:results:projections:control_jpsipi}
\end{figure}

The \PiMuMu mass and \CosThetal distributions of selected \BuJpsiPi candidates are shown in Fig.~\ref{fig:results:projections:control_jpsipi}. 
Excellent agreement is found between the data and  the result of the fit, indicating that the signal efficiency is well described by the corrected simulation.
The \FH and \AFB parameters are found to be consistent with zero, as expected for these decays. 
Similarly, \FH and \AFB determined from the fit to the \BuJpsiK sample are also found to be consistent with zero. 

\begin{figure}[!tbp]
\centering
\includegraphics[width=\linewidth]{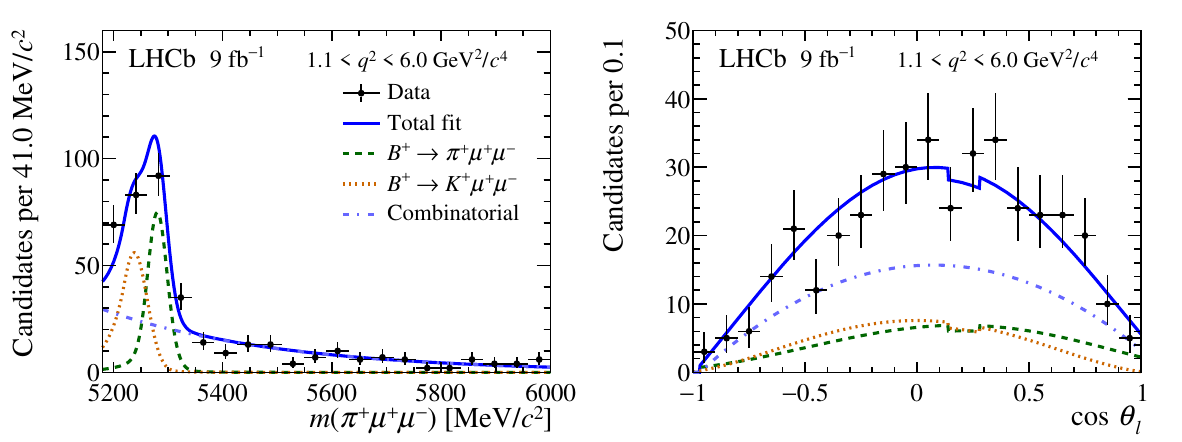} \\ 
\includegraphics[width=\linewidth]{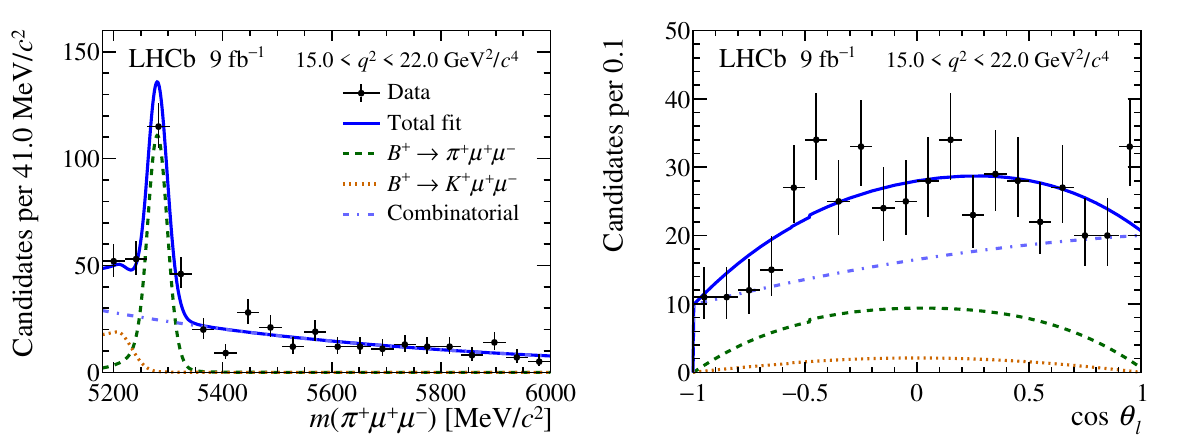} 
\caption{ 
Distribution of (left)~the \PiMuMu mass and (right)~\CosThetal for selected \decay{\Bp}{\pip\mumu} candidates in the range (top)~$1.1 < \qsq < 6.0$\gevgevcccc and (bottom)~$15.0 < \qsq < 22.0$\gevgevcccc.   
The data are superimposed on the result of the fit described in the text.
}
\label{fig:results:projections:signal}
\end{figure}

\begin{figure}[!tbp]
\centering
\includegraphics[width=0.49\linewidth]{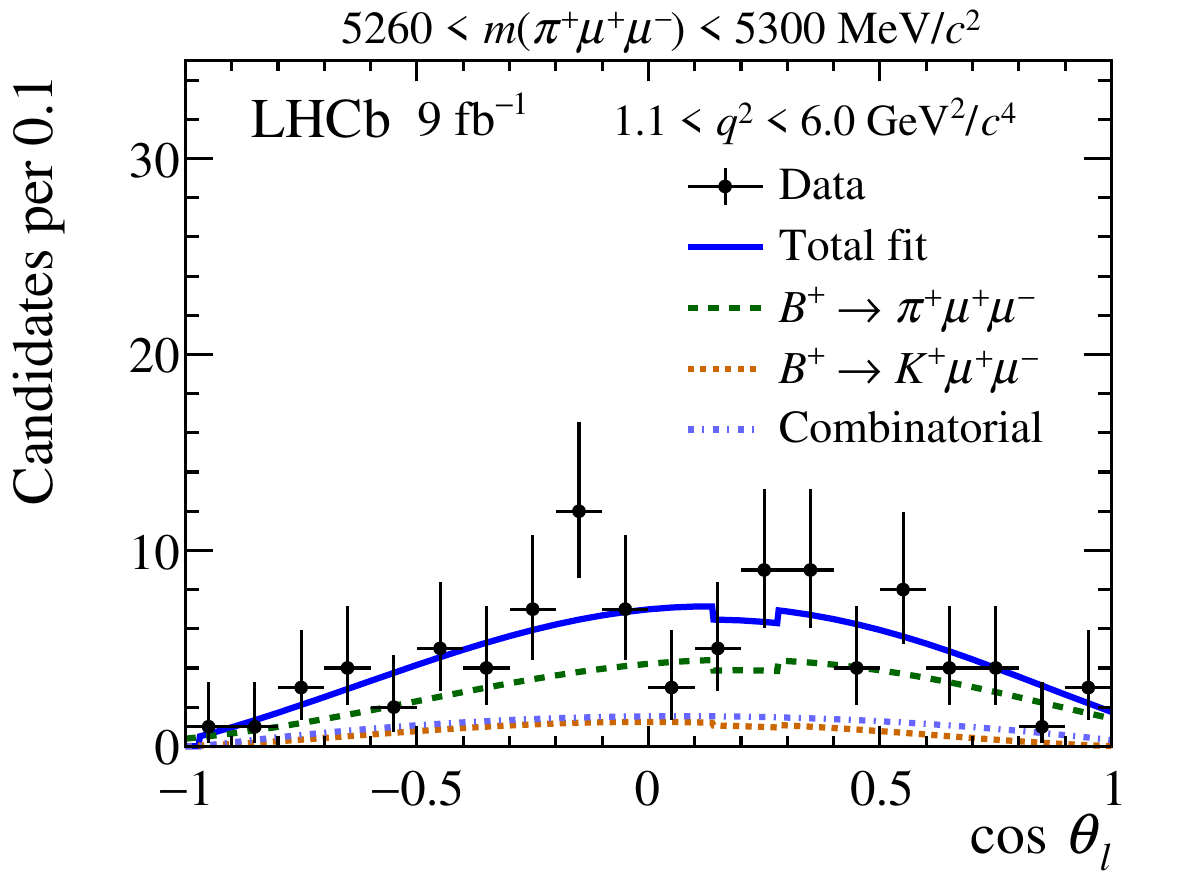}\hfill
\includegraphics[width=0.49\linewidth]{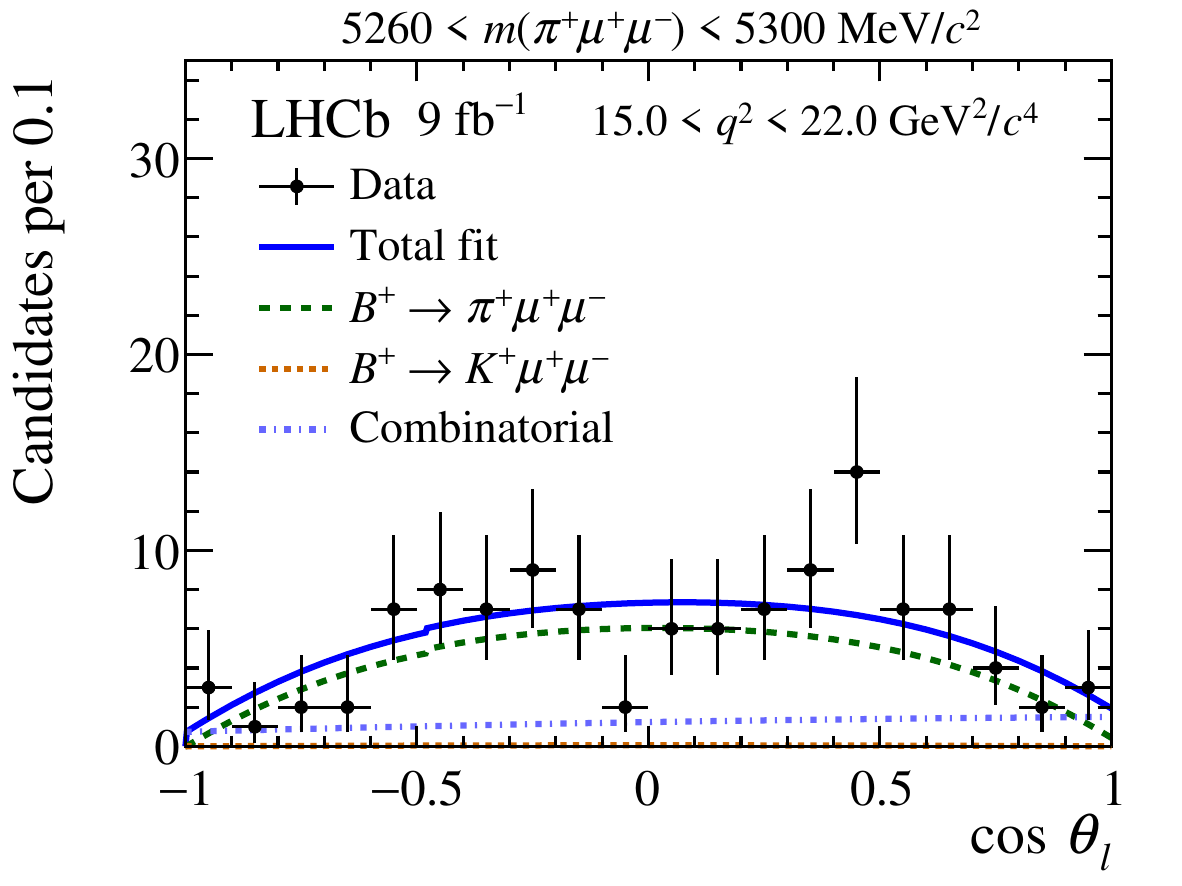} 
\caption{ 
Distribution of \CosThetal for selected \decay{\Bp}{\pip\mumu} candidates in the range (left)~\mbox{$1.1 < \qsq < 6.0$\gevgevcccc} and (right)~$15.0 < \qsq < 22.0$\gevgevcccc.  
The candidates are required to be within the mass interval \mbox{$5260 < m(\PiMuMu) < 5300\mevcc$}, corresponding to a window of approximately twice the \PiMuMu mass resolution around the known \Bp meson mass.
The data are superimposed on the result of the fit described in the text. 
}
\label{fig:results:projections:signal-sigregion}
\end{figure}

Figure~\ref{fig:results:projections:signal} shows the \PiMuMu mass and \CosThetal distributions of the selected candidates in the two signal \qsq\ regions. 
The data are described well by the fit results.
The discontinuity in \CosThetal (visible in the $1.1 < \qsq < 6.0$\gevgevcccc range) arises from the \jpsi veto. 
The observed signal yields are $89\pm13$ and $135\pm15$ in the low- and high-\qsq regions, respectively.
The uncertainties quoted on the yields are purely statistical and are determined by repeating the fit on pseudodata generated by bootstrapping the dataset~\cite{efron:1979}.
The observed signal yields and those of the misidentified \BuKMuMu background are consistent with expectations.   
The expected yields are based on the yields of $B^+\to J/\psi K^+$ decays in data, the branching fractions from Refs.~\cite{LHCb-PAPER-2015-035,PDG2024}, and the relative selection efficiencies and misidentification rates from simulation and control data. 
Figure~\ref{fig:results:projections:signal-sigregion} shows the one-dimensional \CosThetal distributions within the mass interval \mbox{$5260 < m(\PiMuMu) < 5300\mevcc$}, corresponding to a window of approximately twice the \PiMuMu mass resolution around the known \Bp meson mass. 

\begin{figure}[!tb]
\centering
\includegraphics[width=0.49\linewidth]{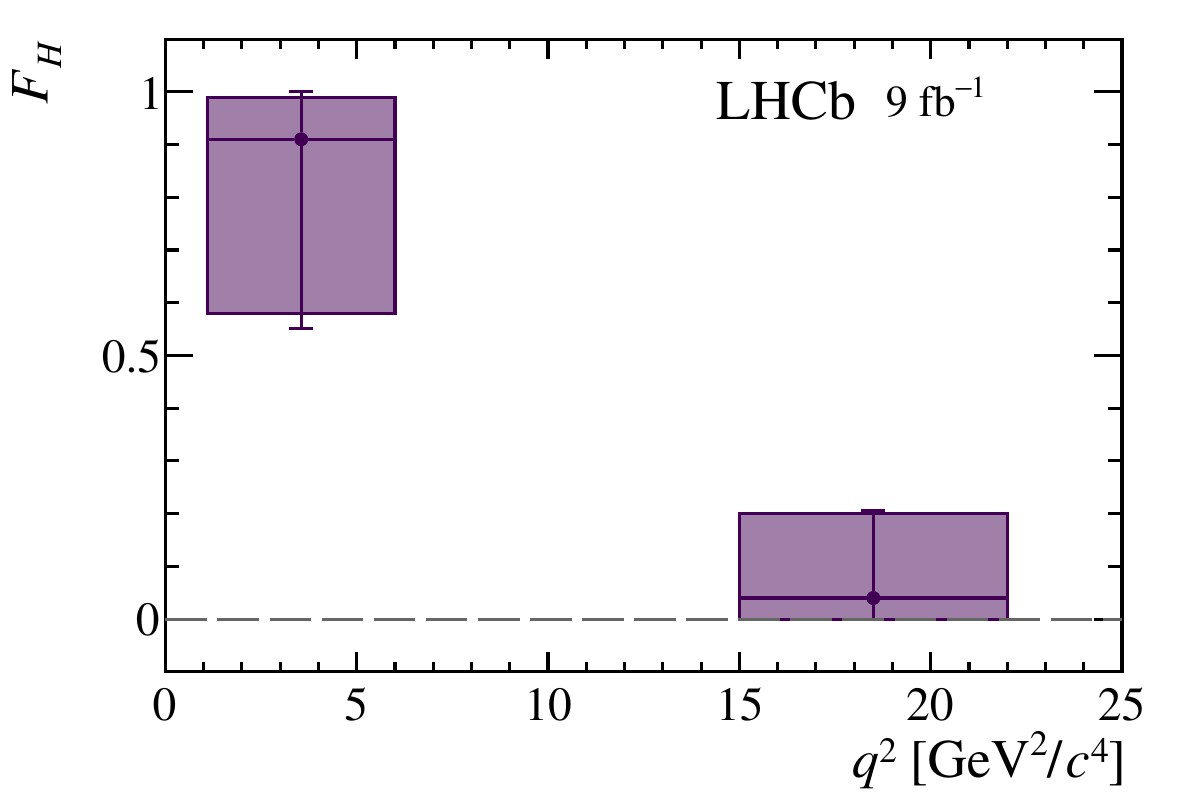}
\includegraphics[width=0.49\linewidth]{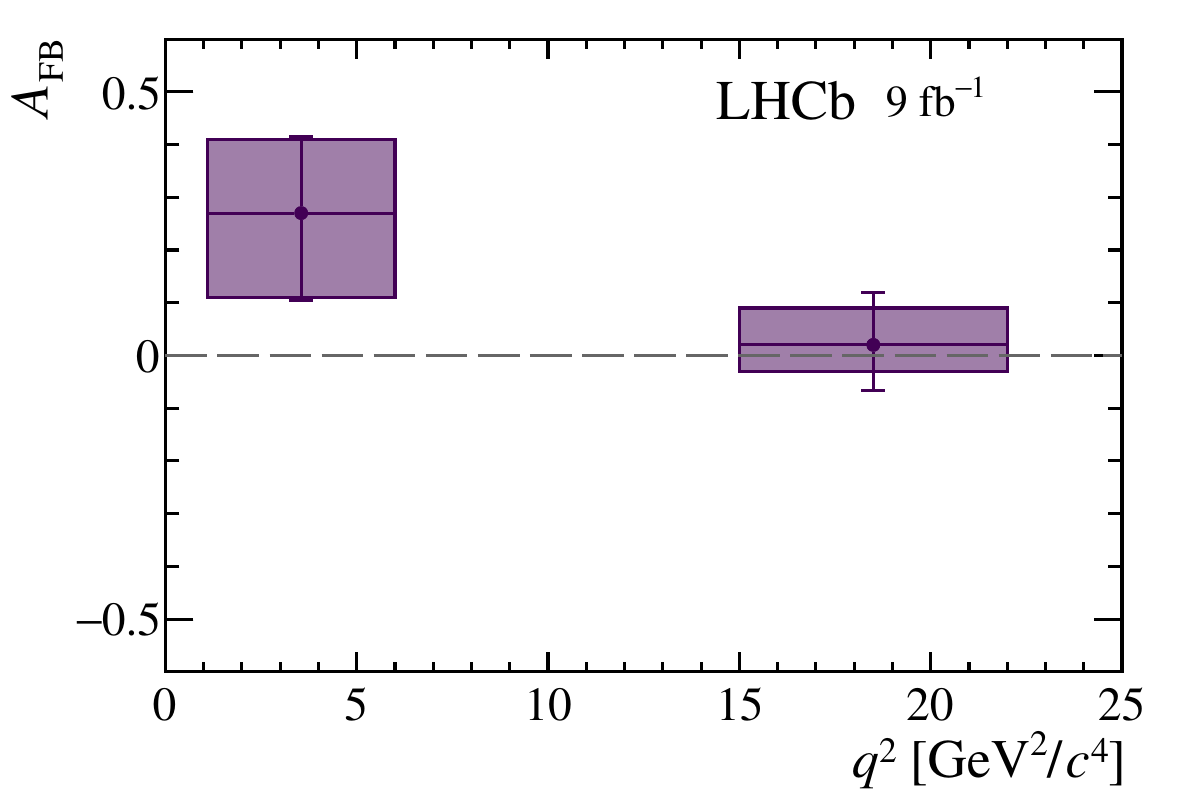} \\
\caption{ 
Obtained results for (left)~\FH and (right)~\AFB in bins of dimuon mass squared, \qsq. 
The markers indicate the best-fit values from the likelihood fit. 
The boxes represent the 68\% CL intervals as determined using the Feldman--Cousins construction.
The outer bars represent the total uncertainties, obtained by adding in quadrature the contribution arising from systematic uncertainties.
}
\label{fig:results:param}
\end{figure}

\begin{figure}[!tb]
\centering
\includegraphics[width=0.48\linewidth]{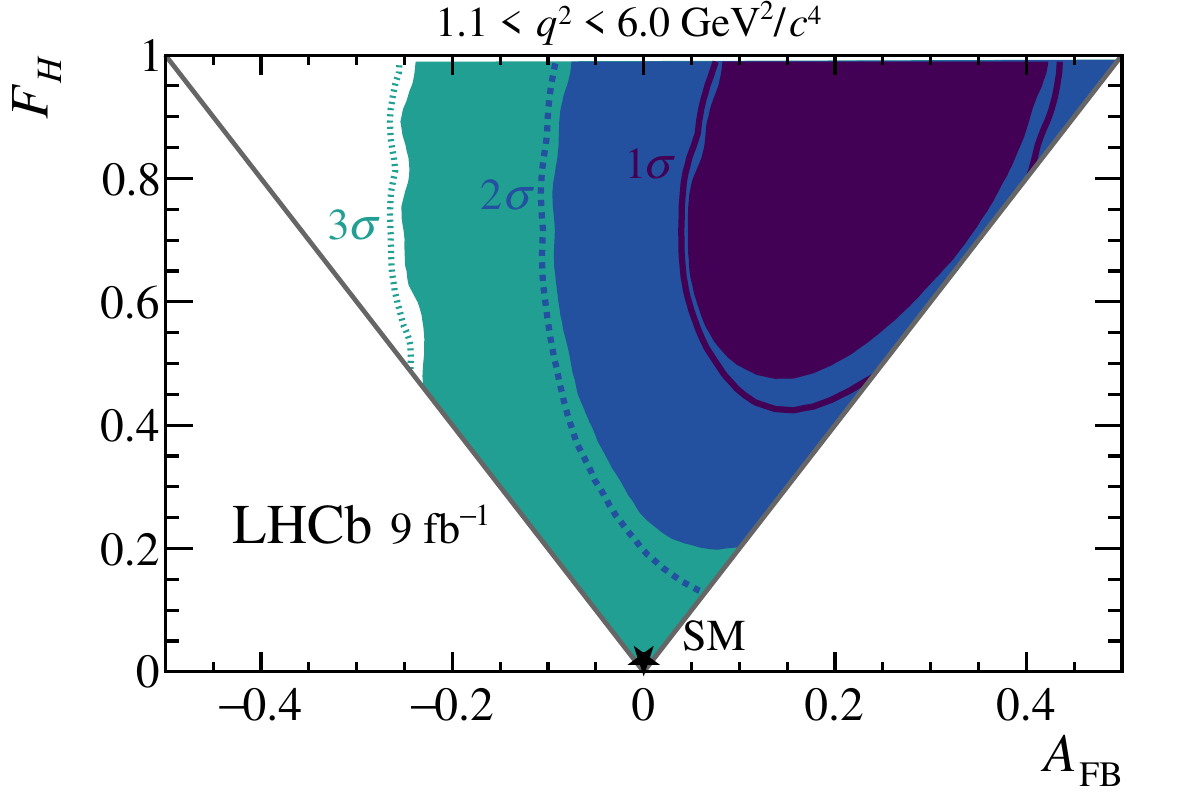}
\includegraphics[width=0.48\linewidth]{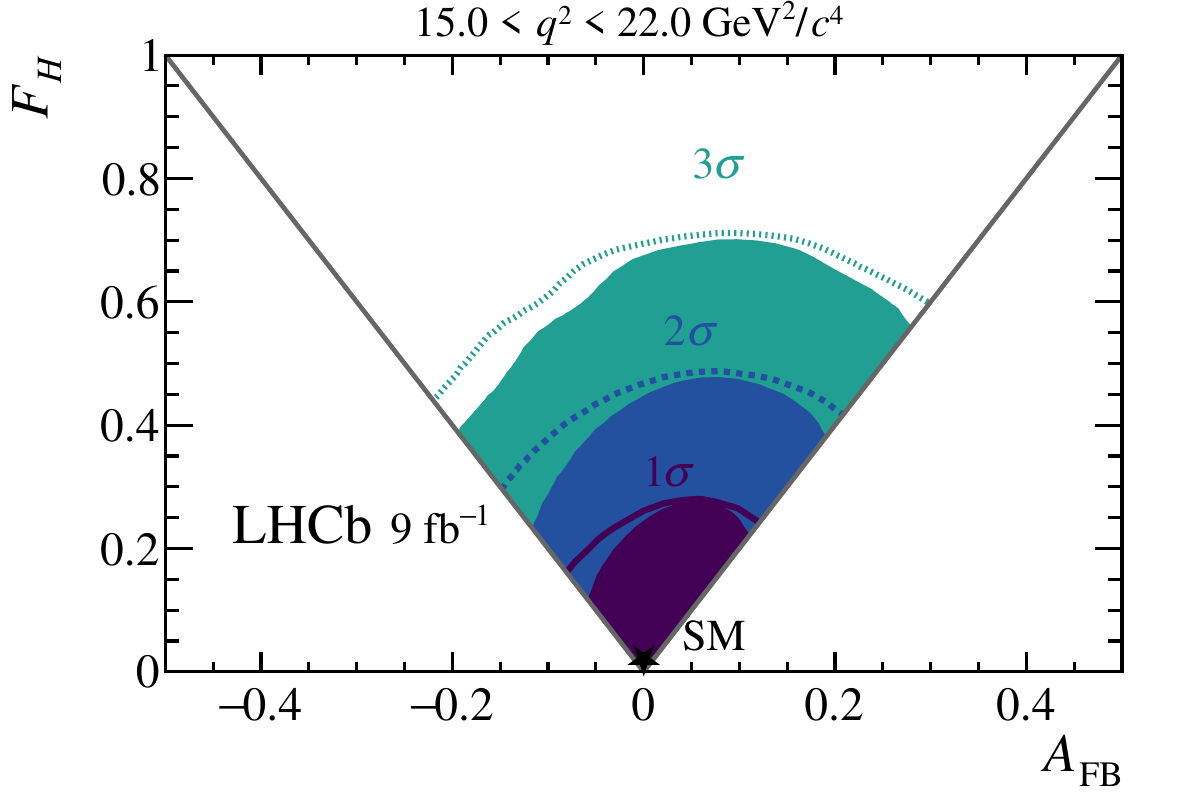} \\
\caption{ 
Two-dimensional CL regions obtained with the Feldman--Cousins procedure for \AFB and \FH in the \qsq\ regions (left)~$1.1 < \qsq < 6.0$\gevgevcccc and (right)~\mbox{$15.0 < \qsq < 22.0$\gevgevcccc}. 
The one-, two-, and three-sigma regions are indicated as different coloured areas, corresponding to the associated statistical intervals. The line contours represent the corresponding confidence levels obtained when systematic uncertainties are combined in quadrature.
The gray lines show the boundary of the physical region in which the probability distribution of the signal remains positive for all values of \CosThetal. 
The data associated with this figure are available on the HEPData repository~\cite{HEPData}.
}
\label{fig:results:twodimensions}
\end{figure}

The values of \AFB and \FH for the two \qsq\ intervals are given in Table~\ref{tab:results} and are shown in Fig.~\ref{fig:results:param}. 
Two-dimensional confidence interval regions for \AFB and \FH are shown in Fig.~\ref{fig:results:twodimensions}. 
At low \qsq, the data show a nonzero \AFB value and a large value of \FH that are not consistent with the SM point within 95\% confidence level (CL), but are consistent within 99\% CL. 
At high \qsq, the data show values of \AFB and \FH that are consistent with the SM point within 68\% CL.

The stability of the results is tested by imposing alternative selection criteria on the output of the boosted-decision-tree background classifier, by applying more stringent particle-identification requirements, and by determining $\FH$ independently of $\AFB$ through an alternative fit to the distribution of the absolute value of \CosThetal. 
All these studies yield results that are consistent with the baseline measurements.

As an additional test, the \CosThetal distribution in data are examined for \BuKMuMu decays reconstructed in the low- and high-\qsq regions. The candidates are selected using the same criteria as for the signal mode, with the exception of the particle-identification requirements on the charged kaon, for which the criteria used in the selection of \BuJpsiK decays are applied. In both the low- and high-\qsq regions, no significant asymmetry is observed for \BuKMuMu decays.

\begin{table}[tb]
    \caption{
    Values of \AFB and \FH obtained from the fit and Feldman--Cousins procedure. 
    The intervals correspond to 68\% confidence level and include statistical uncertainties only. 
    }
    \label{tab:results}
    \centering
    \begin{tabular}{c|cccc}  
    \toprule
     & \FH & \FH interval & \AFB & \AFB interval \\
    \midrule
    $\phantom{1}1.1 < \qsq < \phantom{1}6.0$\gevgevcccc & 
     0.91 & $[0.60, 0.99]$ & $0.27$ & $[\phantom{+}0.11,0.42]$ \\
    $15.0 < \qsq < 22.0$\gevgevcccc & 
     0.04 & $[0.00, 0.20]$ & $0.02$ & $[-0.02,0.09]$ \\
    \bottomrule
    \end{tabular}
\end{table}

\section{Summary} 
\label{sec:Summary} 

The angular distribution of the \BuPiMuMu decay is studied for the first time using a dataset of $pp$ collisions corresponding to 9\invfb of integrated luminosity, collected with the LHCb experiment. 
The angular observables \AFB and \FH, which characterise the angular distribution of the decay, are measured in two intervals of the dimuon mass squared, \qsq: a low-\qsq region defined by $1.1 < \qsq < 6.0\gevgevcccc$ and a high-\qsq region defined by $15.0 < \qsq < 22.0\gevgevcccc$. 
The results are found to be consistent with previous measurements of the corresponding observables in the \BuKMuMu decay~\cite{LHCb-PAPER-2014-007}.
Confidence intervals are constructed according to the Feldman–-Cousins prescription~\cite{Feldman:1997qc}. 
The Standard Model point lies within the 99\% confidence level (CL) interval in the low-\qsq region and within the 68\% CL interval in the high-\qsq region.

%% file: acknowledgements.tex
\section*{Acknowledgements}
%
%
\noindent We express our gratitude to our colleagues in the CERN
accelerator departments for the excellent performance of the LHC. We
thank the technical and administrative staff at the LHCb
institutes.
We acknowledge support from CERN and from the national agencies:
ARC (Australia);
CAPES, CNPq, FAPERJ and FINEP (Brazil); 
MOST and NSFC (China); 
CNRS/IN2P3 and CEA (France);  
BMFTR, DFG and MPG (Germany);
INFN (Italy); 
NWO (Netherlands); 
MNiSW and NCN (Poland); 
MEC/IFA (Romania); 
MICIU and AEI (Spain);
SNSF and SER (Switzerland); 
NASU (Ukraine); 
STFC (United Kingdom); 
DOE NP and NSF (USA).
We acknowledge the computing resources that are provided by ARDC (Australia), 
CBPF (Brazil),
CERN, 
IHEP and LZU (China),
IN2P3 (France), 
KIT and DESY (Germany), 
INFN (Italy), 
SURF (Netherlands),
Polish WLCG (Poland),
IFIN-HH (Romania), 
PIC (Spain), CSCS (Switzerland), 
GridPP (United Kingdom),
and NSF (USA).  
We are indebted to the communities behind the multiple open-source
software packages on which we depend.
Individual groups or members have received support from
RTP (Australia), 
Key Research Program of Frontier Sciences of CAS, CAS PIFI, CAS CCEPP (China); 
Minciencias (Colombia);
EPLANET, Marie Sk\l{}odowska-Curie Actions, ERC and NextGenerationEU (European Union);
A*MIDEX, ANR, IPhU and Labex P2IO, and R\'{e}gion Auvergne-Rh\^{o}ne-Alpes (France);
Alexander-von-Humboldt Foundation (Germany);
ICSC (Italy); 
Severo Ochoa and Mar\'ia de Maeztu Units of Excellence, GVA, XuntaGal, GENCAT, InTalent-Inditex and Prog.~Atracci\'on Talento CM (Spain);
the Leverhulme Trust, the Royal Society and UKRI (United Kingdom).

%% file: Authorship_LHCb-PAPER-2026-015.tex
\centerline
{\large\bf LHCb collaboration}
\begin
{flushleft}
\small
R.~Aaij$^{38}$\lhcborcid{0000-0003-0533-1952},
M. ~Abdelfatah$^{69}$,
A.S.W.~Abdelmotteleb$^{57}$\lhcborcid{0000-0001-7905-0542},
C.~Abellan~Beteta$^{51}$\lhcborcid{0009-0009-0869-6798},
F.~Abudin\'en$^{59}$\lhcborcid{0000-0002-6737-3528},
T.~Ackernley$^{61}$\lhcborcid{0000-0002-5951-3498},
A. A. ~Adefisoye$^{69}$\lhcborcid{0000-0003-2448-1550},
B.~Adeva$^{47}$\lhcborcid{0000-0001-9756-3712},
M.~Adinolfi$^{55}$\lhcborcid{0000-0002-1326-1264},
P.~Adlarson$^{87,42}$\lhcborcid{0000-0001-6280-3851}
% ...

{\footnotesize \it

% ...
$^{38}$Nikhef National Institute for Subatomic Physics, Amsterdam, Netherlands\\
% ...
$^{42}$National Center for Nuclear Research (NCBJ), Warsaw, Poland\\
% ...
$^{47}$Instituto Galego de F{\'\i}sica de Altas Enerx{\'\i}as (IGFAE), Universidade de Santiago de Compostela, Santiago de Compostela, Spain\\
% ...
$^{51}$Physik-Institut, Universit{\"a}t Z{\"u}rich, Z{\"u}rich, Switzerland\\
% ...
$^{55}$H.H. Wills Physics Laboratory, University of Bristol, Bristol, United Kingdom\\
% ...
$^{57}$Department of Physics, University of Warwick, Coventry, United Kingdom\\
% ...
$^{59}$School of Physics and Astronomy, University of Edinburgh, Edinburgh, United Kingdom\\
$^{60}$School of Physics and Astronomy, University of Glasgow, Glasgow, United Kingdom\\
$^{61}$Oliver Lodge Laboratory, University of Liverpool, Liverpool, United Kingdom\\
% ...
$^{69}$Syracuse University, Syracuse, NY, United States\\
% ...
$^{87}$Department of Physics and Astronomy, Uppsala University, Uppsala, Sweden, associated to $^{60}$\\
% ...
}
\end{flushleft}